\documentclass[11pt,twoside,A4]{article} 
\usepackage{times,fancyhdr}
\usepackage[dvips]{graphicx}
\usepackage{latexsym} 
\usepackage[affil-it]{authblk}
\usepackage{mathptm}
\usepackage{amsmath}
\usepackage{amssymb}
\usepackage{setspace}
\usepackage{hyperref}

\usepackage[top=4cm, bottom=4cm, left=4cm, right=4cm]{geometry}

\def\beq{\begin{equation}}
\def\eeq{\end{equation}}
\def\beqn{\begin{eqnarray}}
\def\eeqn{\end{eqnarray}}

\newcommand{\be}{\begin{equation}}
\newcommand{\ee}{\end{equation}}
\newcommand{\bea}{\begin{eqnarray}}
\newcommand{\eea}{\end{eqnarray}}

\begin{document}
 
\title{A Covariant Version of Verlinde's Emergent Gravity}
\author{Sabine Hossenfelder} 
\affil{\small Frankfurt Institute for Advanced Studies\\
Ruth-Moufang-Str. 1,
D-60438 Frankfurt am Main, Germany
}
\date{}
\maketitle
\begin{abstract}
A generally covariant version of Erik Verlinde's emergent gravity
model is proposed. The Lagrangian constructed here allows 
an improved interpretation of the underlying mechanism. It suggests that de-Sitter
space is filled with a vector-field that couples to baryonic matter and,
by dragging on it, creates an effect similar to dark matter. We solve the
covariant equation of motion in the background of a Schwarzschild
space-time and obtain correction terms to the non-covariant expression.
Furthermore, we demonstrate that the vector field can also mimic
dark energy. 

\end{abstract}

\section{Introduction}

In a recent work, Erik Verlinde \cite{Verlinde:2016toy} put forward a new approach to emergent
gravity. According to this hypothesis, dark matter is not made of unknown, weakly interacting
particles. Instead, it is a relic of the 
fundamental degrees of freedom of a quantized space-time which make themselves
noticeable as a modification of general relativity. This idea is interesting 
because it allows us to connect the quantum description of space-time with
observable consequences.  

However, the so-far provided framework raises many questions. In
particular, the original version does not have a Lagrangian formulation and is applicable only
to spherically symmetric systems in certain limits. 

The purpose of this
present work is to 
identify a Lagrangian for the idea of \cite{Verlinde:2016toy} in the hope that
it serves to catalyze future work. We will also study the spherically symmetric
and spatially homogeneous case in the fully covariant framework.

Throughout this paper, we use units in which $c=\hbar=1$ and $G=m_{\rm p}^2$, where $m_{\rm p}$ is the Planck
mass. Small Greek indices run from 0 to 3 and 
denote space-time dimensions; small Latin indices run from 1 to 3 and denote
spatial dimensions only. Bold-faced quantities denote tensors with indices suppressed. 
The signature of the metric is $(-,+,+,+)$.

\section{Summary of previous works}

For the benefit of the reader, we first briefly summarize the main findings
of  \cite{Verlinde:2016toy}. The new idea builds on earlier work in which
it was demonstrated that general relativity can be obtained from
a thermodynamic approach \cite{Jacobson:1995ab,Cai:2005ra,Padmanabhan:2009vy,Verlinde:2010hp}.  For this, one assumes that the entropy contained in 
a volume scales with the surface area of that volume. 

It was also previously noted  that {\sc MO}dified
Newtonian Dynamics (hereafter {\sc MOND}) \cite{Milgrom:1983zz,Milgrom:1983ca} has a characteristic acceleration
scale ($a_0$) which -- by its order of magnitude -- seems related to the temperature
of the (approximate) de-Sitter space that we live in, ie the cosmological
constant, by $a_0^2 = \Lambda = 1/L^2$, where $L$ is
the (current) Hubble radius and $\Lambda$ is the cosmological constant . 

The key idea of \cite{Verlinde:2016toy} is to alter the entropy used to
derive general relativity so that one instead obtains a modified theory
of gravity which mimics dark matter. This is done by adding a volume-dependent term to the
entropy. This term is chosen so as to reproduce the entropy expected for de-Sitter space
with the observed cosmological constant. 

It is useful to think
of the additional entropy as corresponding to a type of field 
that fills the universe. This interpretation should in any case be possible because
a modification of gravity can always be reformulated as a non-standard
contribution to the source-term of gravity. So, effectively, the additional entropy
adds additional stuff, with properties to be specified. Ideally, this description would
emerge as the low-energy limit of a more fundamental theory.

Since we
expect the additional stuff to play a role for both dark matter and dark energy without
being either, 
we will refer to it as the `imposter field.' 

According to  \cite{Verlinde:2016toy}, the behavior of the imposter is quantified
by a vector-field, ${\bf u}$, which is interpreted as the displacement-vector
of an elastic medium. It is then argued that the entropy of the field is reduced in
the vicinity of matter, and this reduction in turn 
corresponds to a decrease of volume occupied by the field. This, then, generates 
a  force pointing towards the matter,
which can be felt by other matter. 

We here do not wish to sign up
to the interpretation of the medium being elastic and ${\bf u}$ being
the displacement vector, which is why we chose the more neutral
nomenclature of the imposter field.

The additional force
created by the imposter can be reinterpreted as being due
to additional matter. It is
then shown in  \cite{Verlinde:2016toy} that -- at least
in first estimates -- it has the
right behavior to account for the observed dark matter
in galaxies. 

A great benefit of this approach is that the entropy of the imposter field cannot become smaller 
than zero. This means if the density of matter is too large, the entropy-decrease
bottoms out and one is left only with the area-scaling part of the entropy.
With this approach one can therefore explain why one does not
observe {\sc MOND}-like modifications on solar-system scales.

Since the additional entropy is related to the cosmological
constant, \cite{Verlinde:2016toy} reproduces a suitable
acceleration scale that can be identified with the {\sc MOND}
scale. We note that \cite{Verlinde:2016toy} starts from the
assumption that space-time is (approximately) asymptotically de-Sitter
space. Hence, one obtains dark matter as a side-effect of
dark energy, but one has to put in dark energy by hand.

Whether this approach can also account for
other observational evidence for dark matter, notably the
power spectrum of the temperature fluctuations in the
cosmic microwave background, is an open
question. The formalism offered so far in \cite{Verlinde:2016toy}
does not lend itself to the necessary analysis required for
cosmological perturbation theory.

This new approach to emergent gravity is promising. It 
is, however, still immature and
has some shortcomings.

First, the approach does not have a
Lagrangian. A Lagrangian would be preferable to make sure the equations are 
internally consistent. It would also make it easier to tie this idea to other models, and identify
differences and similarities. 

Indeed, one may suspect that the equations in \cite{Verlinde:2016toy}  are
internally inconsistent as it stands just because there
are too many of them. Eg, the conservation of ${\bf \sigma}$ in Eq.\ (7.21) \cite{Verlinde:2016toy} gives a differential
equation for ${\bf u}$ by use of Eqs.\ (6.1) \cite{Verlinde:2016toy} and (7.20) \cite{Verlinde:2016toy}. Eq.\ (7.36) \cite{Verlinde:2016toy} also gives a differential equation for {\bf u}, and the
two are generically not identical. 

Moreover, the definition
for ${\bf u}$ in Eq.\ (4.38) \cite{Verlinde:2016toy} relates the displacement vector with
the Newtonian potential in a way that does not fit with either 
equation. A further problem is that the force that is derived by Eq. (6.17) \cite{Verlinde:2016toy} is not the same as
the force one obtains from (7.36) \cite{Verlinde:2016toy}. This indicates that the unclear definition of physical
quantities is not merely a matter of interpretation.

Second, the expressions derived in \cite{Verlinde:2016toy} are not covariant.
This is clear already by selecting a
spherically symmetric configuration, but even without spherical symmetry 
it is not obvious how to generalize this covariantly. Furthermore, it remains somewhat mysterious what
happens with the covariant-looking derivatives that appear throughout the 
paper but later are not actually treated as covariant derivatives.

Third, the approach in \cite{Verlinde:2016toy} makes some assumptions that
are probably over-idealized limits. For example, if we think of (normal) matter
as inducing a phase-transition in the imposter field, then the imposter's entropy should not
simply drop to zero at some point. It might drop discontinuously to an unnoticeably 
small remainder, and zero might be a good approximation. But it seems unphysical 
to assume that it would no longer have any gradient. Likewise, the
medium should not be entirely incompressible, or perfectly elastic,
or totally lack local excitations. 

All of these points could be addressed with an effective Lagrangian.

\section{Constructing a Lagrangian}

The advantage of working with Lagrangians is that symmetry-requirements and
dimensional analysis are often sufficient to at least arrive at a working example.
This is what we will do here.

Let us begin by identifying the mass-dimensions of the unfamiliar, new quantities
introduced in  \cite{Verlinde:2016toy}. The surface mass density, $\Sigma$,
has a dimension of mass per area, ie mass dimension three. The displacement-vector, ${\bf u}$, has the 
dimension of an inverse mass, so that its derivatives are dimensionless. It is useful
to introduce the strain tensor
\beqn
\epsilon_{\mu \nu} = \nabla_\mu u_\nu + \nabla_\nu u_\mu~,
\eeqn
and the dimensionless scalar
\beqn
\phi := \frac{\sqrt{- u^\alpha u_\alpha}}{L}~,  \label{pot}
\eeqn
which can be identified with the gravitational potential associated with the imposter field. The derivative $\nabla$ here is 
assumed to be the usual,
torsion-free connection compatible with the metric ${\bf g}$. 
For convenience, let us also introduce the abbreviations
\beqn
u := \sqrt{-u^\nu u_\nu} ~,~ \epsilon := \epsilon^{\kappa}_{\;\; \kappa}  ~.
\eeqn
\cite{Verlinde:2016toy} further uses the field ${\bf n}$ which the normalized ${\bf u}$ and dimensionless.

To obtain the Lagrangian, we start with a  look at equation (7.37) \cite{Verlinde:2016toy}. We see that it has the following
scaling behavior: On the left side we have an expression quadratic in the
derivatives of ${\bf u}$ -- sketchily, 
$ (\nabla {\bf u})^2$ -- which is good. The right side, however, which 
should describe normal (baryonic) matter is not
in a form expressed by familiar quantities due to the appearance of ${\bf n}$. 

To remedy this problem, we take a covariant derivative on both sides. 
Then, on the left side we have an expression of the form $\nabla (\nabla {\bf u})^2)$.
On the right side we now have the
second derivative of the potential, but this is just the source. This means that,
sketchily again, the right side has the form of ${\bf Tn}$, where ${\bf T}$ is
the stress-energy-tensor of the normal matter. Let us further conjecture that these are the 
four equations of motion
for the imposter field ${\bf u}$.

This lets us guess that the Lagrangian for the free imposter field is of the form ${\cal L} \sim (\chi)^{3/2}$, where $\chi$ 
is the kinetic term. The kinetic term for a vector field can in general have three components, with
different contractions of the indices:
\beqn
\chi = \alpha (\nabla_\nu u^\nu ) (\nabla_\kappa u^\kappa ) + \beta  (\nabla_\nu u_\kappa ) (\nabla^\nu u^\kappa ) + \gamma  (\nabla_\nu u_\kappa ) (\nabla^\kappa u^\nu ) ~. \label{chi}
\eeqn
The parameter combination $\alpha=0, \gamma = - \beta$ is the only possible combination for a gauged vector field, but we
have no reason to assume that the imposter field is some kind of gauge-boson. Indeed, since it is the symmetric
combination of derivatives that appears in the strain-tensor, the case with $\gamma = \beta$ makes more sense.

We can moreover extract from Eq.\ (6.5) \cite{Verlinde:2016toy} that the matter in the
background of the imposter field feels an effective spatial metric of
the form
\beqn
{\widetilde h}_{ij} = h_{ij} - u_i n_j /L~. \label{geff}
\eeqn
Where ${\bf n} \parallel {\bf u}$ and $n^\alpha n_\alpha = -1$. We have allowed
ourselves some freedom here, since (6.5) \cite{Verlinde:2016toy} does not actually use a general
spatial metric, $h_{ij}$, but merely a flat one, $\delta_{ij}$. 

We note in the
passing that this form of the metric is not so unlike the effective acoustic metric \cite{Barcelo:2005fc}, except
for the somewhat unusual convention to normalize one of the vectors to one, rather
than requiring both to have the same normalization. 
The expression (\ref{geff}) leads us to expect that normal matter couples not as usual with
${\bf gT}$ but with $({\bf g-un}/L){\bf T}$\footnote{A more general way to do it would be a type of minimal coupling by replacing ${\bf g}$ with
${\bf g-un}$ in the usual matter Lagrangian. The difference is a term proportional to ${\cal L}$, which
however in the examples we are studying here does not make a contribution. Therefore, this ambiguity need
not bother us for the purposes of this paper but should be reconsidered for more general situations.}. 

If we now factor in the relevant constants, we obtain following total 
Lagrangian:
\beqn
{\cal L}_{\rm tot} &=& m_{\rm p}^2 {\cal R} + {\cal L}_{\rm M} + {\cal L}_{\rm int} + {\cal L}_{\theta} ~, \label{lag} \\ 
{\cal L}_{\rm int} &=& - \frac{1}{L} u^\mu n^\nu T_{\mu \nu} = - \frac{u^\mu u^\nu}{L u} T_{\mu \nu} ~, \nonumber \\
{\cal L}_{\theta} &=& \frac{m_{\rm p}^2}{L^2} \chi^{3/2}  - \frac{\lambda^2 m_{\rm p}^2}{L^4} u_\kappa u^\kappa  ~, \nonumber
\eeqn
where ${\cal L}_{\rm M}$ is the Lagrangian of the normal matter. As usual, we have
\beqn
\frac{\delta {\cal L}_{\rm M}}{\delta g^{\mu \nu }}~ = - \frac{1}{2} \left( T_{\mu \nu} - g_{\mu \nu} {\cal L}_{\rm M} \right)	~,
\eeqn
where $T_{\alpha \kappa}$ is the stress-energy tensor of the normal matter. 

The Lagrangian could also have a self-interaction
term for ${\bf u}$, but since we do not presently have enough information to
construct one, we omit it. 

Since the
dimension of the imposter field and its normalization is so unusual, it is 
not all that clear which dimension the mass-term should have. However, for
the present purposes it 
turns out not to matter all that much which power of ${\bf u}$ appears in this term. We
need a mass-term however because otherwise we expect the vector ${\bf u}$ to
be light-like, clashing with its normalization.

In total, we have four dimensionless parameters in the Lagrangian, $\alpha,\beta,\gamma,$ and $\lambda$. This
seems a lot, given that we do not have all that many constraints on the model. For
the rest of the paper, we will therefore use the combination $\alpha=4/3, \beta = \gamma = -1/2$, so that
\beqn
\chi= - \frac{1}{4} \epsilon_{\mu \nu} \epsilon^{\mu\nu} + \frac{1}{3} \epsilon^2 ~.
\eeqn

A brief explanation for this particular choice can be found in the appendix.

\section{Reproducing the non-relativistic limit}

With the Lagrangian (\ref{lag}), the equations of motion for ${\bf u}$ work out to be
\beqn
-\frac{3m_p^2}{2L} \nabla^\nu \left( \left( - \epsilon_{\nu \mu}  + \frac{4}{3} \epsilon \right) \sqrt{\chi} \right) =  
n^\nu \left( 2 \delta_\mu^{\;\;\kappa}   -  n_\mu n^\kappa \right)  T_{\kappa \nu} - 2 \frac{\lambda^2 m_{\rm p}^2}{L^3} u_\mu    ~.  \label{eom}
\eeqn
This looks complicated, but some simplifying
assumptions will help. 

First, we assume that we can neglect the terms stemming from the
mass of the imposter field. (A mass of the vector field did not appear 
in \cite{Verlinde:2016toy}.) We will justify this below. 

Second, we assume that we work
in or close by the restframe of the imposter field, so that ${\bf n} \approx (-1, \vec 0)$. Third, as in \cite{Verlinde:2016toy},
we assume that  the solution we search for is both time-independent and spherically symmetric, which means the functions we want to solve
for depend only on the radial coordinate, $r$.  Fourth, we assume that we are dealing with a pressureless
fluid, so that the only nonzero component of the stress-energy-tensor is
$T_{00} = \rho$.  Finally, as in \cite{Verlinde:2016toy}, we approximate the covariant derivatives with
partial derivatives which disregards space-time curvature.

With these assumptions, we see that the right side of Eq.\ (\ref{eom}) simplifies to just $-\rho$ for $\mu = 0$ and
zero for the other components. We then have only one equation for the absolute value of ${\bf u}$ or the potential, $\phi$, from Eq.\ (\ref{pot}), respectively, which is
\beqn
\frac{3m_p^2 L}{2} \partial_r (\partial_r \phi )^2 = \rho~. \label{eomsimple}
\eeqn

For the case of a point-mass, $M$, the source
is a delta-function. One integrates it and gets back Eq.\ (7.37) \cite{Verlinde:2016toy}, from
which one obtains
\beqn
\phi(r) = \sqrt{\frac{M}{L}}\frac{1}{m_{\rm p}} \ln \left( \frac{ r m_{\rm p}^2}{M} \right) + C_1~, \label{mondphi}
\eeqn
where $C_1$ is a constant of integration.
Such a logarithmic potential leads to a $1/r$-force, which results in flat rotation curves. Hence, this
solution has a chance to explain observations. 

With this potential, we can see that the requirement $\epsilon^2 \gg u^2/L^2$ is fulfilled so long as $L^2 \gg r^2$,
ie unless we work on cosmological scales. Thus, for galaxies and galaxy clusters it is a consistent approximation
to neglect the mass of the imposter field.

Admittedly, however, it is not clear under which circumstances all these approximations are valid,
but at least we now know which approximations are
being made. Especially
the use of partial instead of a covariant derivatives is not easy to justify. While a correction
term to the partial derivative stemming from the non-trivial background might be negligible
relative to the leading order of $\phi$, it is not obvious that this correction term
would also be negligible relative to the usual Newtonian potential.

The integration constant in the above expression should be fixed by the asymptotic limit. If we
require that the $00$-component of $({\bf g-un}/L)$ still asymptotes to $-1$ for $r\to \infty$, then $C_1=0$.
Taking this limit, however, is problematic because several of the
above made assumptions no longer apply. It is therefore not clear how to fix the integration
constant. What we are missing here is a boundary condition that will match this potential to
the embedding in the cosmological background, a problem we will come back to later.

It should
be pointed out that the differential equation (\ref{mondphi}) for the potential was proposed long ago for 
(non-relativistic) MOND \cite{MOND1}, and a relativistic completion which
also employs the power $3/2$ for the kinetic term in the Lagrangian was discussesd in \cite{Bekenstein:2004ne}. The 
key difference is that in the present work the fundamental field is a vector
field, and the previously discussed equation is merely that for its absolute value which -- assuming a static, spherically
symmetric case --  is the only remaining equation of four. The role of massive vector fields for
cosmology, on the other hand, was studied in \cite{DeFelice:2016yws}, but in this case not with
the power $3/2$ in the kinetic term. We hence see that the approach considered here combines these
two ingredients. 

It is further interesting
that, at least so far, this model seems to work without the need to add a scalar field, as is the
case with modified gravity of the scalar-vector-tensor type \cite{Moffat:2005si}.

\section{Learning from the Lagrangian}

This Lagrangian, (\ref{lag})  results in various subtle 
differences to the proposal of  \cite{Verlinde:2016toy}.

The first difference to Verlinde's approach is that the extra terms in
the Lagrangian also make contributions to the field equations. Just
setting the matter source to zero does not generally imply that the 
imposter field ${\bf u}$ identically vanishes. One hence is not
done with solving the equations of motion for ${\bf u}$, one also
has to calculate the field equations with that source added, which
was not done in  \cite{Verlinde:2016toy}, and take into account
that the motion of matter is affected by both, the curvature
of space-time plus the additional field. 

Indeed, this must be so for physical reasons. If the imposter can
transfer energy to normal matter by exerting a force, then stress-energy-conservation
(realized by the Bianchi-identities) demands that the imposter itself
also acts as a source for gravity. It is apparent however that the stress-energy-tensor
related to the imposter (which can be derived from the Lagrangian) 
is not the stress-energy-tensor $\sigma$ from \cite{Verlinde:2016toy}, which
is the reason why the two ways of calculating the force do not agree in   \cite{Verlinde:2016toy}.
 
Concretely, this means in Eq.\ (\ref{eom}) one cannot simply assume that the source
on the right side itself fulfils the usual Newtonian limit because the
equation of motion for the source also couples to ${\bf u}$. In fact, it better
should not, because that would not make any sense -- one wants the matter
to feel the additional force.

However, we can also see that the actual stress-energy tensor $\widetilde T$ belonging to
the imposter field is of the order $m_p^2 (\nabla u)^3 / L^2$, so that
for the $\phi$ from Eq.\ (\ref{mondphi}) it is of the order 
\beqn
\widetilde T \sim \sqrt{\frac{M}{L}} \frac{M}{m_{\rm p}} \frac{1}{r^3}~.
\eeqn
This means that for the masses under consideration here (galaxies or galaxy clusters), this contribution
is negligible compared to that of normal baryonic matter. It is therefore justified
to neglect it. 

This might potentially be confusing because we are used to dark matter being 
the dominant component on galactic scales. It is therefore worthwhile to point out the distinction to be made
here: The additional force that is experienced by normal matter in this model
does {\it not} stem from the gravitational pull of the additional field ${\bf u}$. 
Instead, the additional force stems
from the direct interaction of the imposter with normal matter (as per the interaction term in
the Lagrangian). The above estimate shows that, compared to that direct
force, on galactic scales the gravitational pull from the additional matter is negligible. 
Another way to say this is that the apparent dark matter distribution
which is introduced in \cite{Verlinde:2016toy} is not the stress-energy associated
with the displacement-vector-field. 

This suggests the following interpretation for the mechanism proposed
in \cite{Verlinde:2016toy}. In the vicinity of matter, the imposter field can
condense. This reduces the volume of
the imposter, causing the surrounding to push in, thereby dragging
on the matter and creating an additional force. If the potential is
strong enough, the imposter can enter
a superfluid phase. In the superfluid phase, the entropy is dramatically
reduced and the friction with normal matter is basically zero. This
means there won't be any modification of general relativity. To match
with the scaling proposed in \cite{Verlinde:2016toy}, the transition
should take place approximately when $\epsilon^2 \sim 1$. 

This means that we should think of the previously proposed Lagrangian (\ref{lag})
as describing the non-superfluid phase. Admittedly, an accurate 
picture of the situation must be more complicated than that because
the fluid in general should be a mixture of both fluid and superfluid
components and not just suddenly turn entirely superfluid at some
point, but such a treatment is beyond the aim of this current work.
Also, it would make more sense to use an assumption of
(approximate) equilibrium rather than just assuming a time-independent
solution, for example using the virial theorem \cite{CD}.

A second benefit of using the Lagrangian is making a connection to other
variants of modified gravity. For example it is interesting to note that the
 strange-seeming power of $2/3$ in the Lagrangian was also used
in \cite{Berezhiani:2015bqa}. In \cite{Berezhiani:2015bqa}, the authors showed that
a superfluid with this type of Lagrangian can reproduce {\sc MOND}-like behavior in
the superfluid phase. But -- much like Verlinde's emergent gravity -- has the benefit
of displaying a different behavior on short distances, where a large gradient
near sources of high energy-density (like, eg, stars) renders the superfluid
approximation invalid. The approach in  \cite{Berezhiani:2015bqa}, however, uses
a scalar field and differs in the interpretation from what we proposed here.

\subsection{The covariant, spherically-symmetric case}

We will now approach the question how a solution to the generally covariant
equations of motion can be found in the static, spherically-symmetric case. 

Since we
have previously seen that it is consistent to neglect the mass-term on galactic
scales, we will for now set $\lambda =0$. We will also, for now, not take into
account the backreaction of the imposter field on the metric. That is,
we will merely solve the equations of motion for ${\bf u}$ in the background
of a Schwarzschild-metric, instead of finding a self-consistent solution for
the background that also takes into account the stress-energy of ${\bf u}$.
In the previous section we have estimated that this, too, is a consistent
approximation.

We place the field in the background of the Schwarzschild-metric in
the usual coordinates
\beqn
ds^2 = - \gamma(r)  {\rm d}t^2 + \frac{1}{\gamma(r)} dr^2 + r^2 {\rm d} \theta^2 + r^2 \sin(\theta)^2 {\rm d}\phi^2~, \label{ssm}
\eeqn
where 
\beqn
\gamma(r) = 1 - \frac{2M}{m_{\rm p}^2 r}~.
\eeqn
The kinetic term of ${\bf u}$ then takes the form
\beqn
\chi = \frac{1}{2} \frac{1}{r^4 \gamma(r)^2} \left(  r^2 \gamma (r) \partial_r \phi - 2 M \phi(r) \right)^2~. \label{chirad}
\eeqn

The diagonal elements of $\nabla {\bf u}$ vanish, so that $\epsilon = 0$, and 
the equations of motion reduce to 
\beqn
\nabla_\nu \left( \sqrt{\chi} \epsilon^{\nu}_{\;\; \mu} \right) = 0 \quad {\mbox{for}}~\quad r \neq 0~.
\eeqn 
We make the ansatz ${\bf u} = (\phi(r) L ,\vec 0)$ and note that, in this case, the strain-tensor takes the form
\beqn
\epsilon_{\mu \nu} \sim  \sqrt{\chi}
\left(
\begin{array}{cccc}
0 & 1 & 0 & 0 \\
1 & 0 & 0 & 0 \\
0 & 0 & 0 & 0 \\
0 & 0 & 0 & 0 \\
\end{array}
\right)
\eeqn
This means that the equations of motion will be fulfilled if and only if
\beqn
\chi \propto \frac{1}{r(r-2M)}~.
\eeqn
With this, Eq.\ (\ref{chirad}) can readily be integrated to
\beqn
\phi(r) \propto  2 \sqrt{\gamma(r)} - \gamma(r) \ln \left( - 1 + \frac{r m_{\rm p}^2}{M} \left( 1+  \sqrt{\gamma(r)} \right)\right)  + C_2 \gamma(r) ~, \label{phiSSM}
\eeqn
where $C_2$ is a constant of integration.

For $rm_{\rm p}^2 \gg M$, the potential (\ref{phiSSM}) can be approximated with
\beqn
\phi(r) &\propto& \gamma(r) \left( - C_3 + \ln \left( \frac{r m_{\rm p}^2}{M} \right)   \right)
+ \left( \frac{3M}{m_{\rm p}^2 r} \right)   \nonumber \\ 
&-& \frac{9}{4} \left( \frac{M}{m_{\rm p}^2 r} \right)^2
- \frac{5}{3} \left( \frac{M}{m_{\rm p}^2 r} \right)^3 + {\cal O} \left( \left( \frac{M}{m_{\rm p}^2 r} \right)^4\right) ~,
\eeqn
where $C_3 = -2 + \ln(2) + C_2$. We want to remind the reader that in addition we are also still in the limit $r\ll L$.

The pre-factor of the potential is not fixed here, because we have not taken into account that there is a delta-function peaked at $r=0$, the
integral over which has to give back the total mass $M$. This is also the reason why $L$, which comes from the coupling to
the source, does not appear in Eq.\ (\ref{phiSSM}). 

We take the point-source into account by choosing the prefactor 
to match the non-relativistic limit, Eq.\ (\ref{mondphi}) which gives
\beqn
\phi(r) = \sqrt{\frac{M}{L}} \frac{1}{m_{\rm p}^2} \left[ \gamma \left( -C_3+ \ln \left(  \frac{r m_{\rm p}^2}{M} \right) \right) + \left( \frac{3M}{m_{\rm p}^2 r} \right) 
+ {\cal O} \left( \left( \frac{M}{m_{\rm p}^2 r} \right)^2\right) \right] ~. \label{phiapprox}
\eeqn
As in the non-relativistic case, we have no good way to fix the integration
constant because we are missing a boundary condition or initial value. The
constant should be determined by taking the limit $r\to \infty$, but this limit
is not covered by our approximation.

This integration constant is important because we see that, in the covariant
treatment, it is multiplied with $\gamma(r)$, which means that $\phi$
obtains a term that goes with $1/r$ whose prefactor depends on the
unknown constant. Without fixing the integration constant, it is thus not
clear whether this term is negligible relative to the usual Newtonian
potential. 

To solve this problem, we would have to find a solution for the coupled system
of field equations and equations of motion, with both backreaction and mass
taken into account. Lacking this solution, the integration constant must be
treated as a free parameter. 

Moreover, to obtain an expression for the apparent dark matter density expected for galaxies, 
one would have to find a way to solve the equation of motion in the background stemming from
the density of baryonic matter in galaxies or galaxy clusters, coupled to the source. 
While it is straight-forward now to write down the equations, solving them is best done 
numerically (especially if not spherically symmetric) and beyond the scope of this 
present work.

\subsection{Cosmology}

The third benefit of having a Lagrangian is that we can ask what is the effect of $\bf{u}$ on 
cosmological scales. Let us look at what happens if we have a universe
filled with only the imposter field and no normal matter, ie ${\bf T} \equiv 0$. For
this, we first derive the source-term stemming from ${\bf u}$. Its stress-energy
tensor is\footnote{A factor is wrong in this and some of the following equations.
Conclusions are not affected, but please refer to \cite{Dai:2017guq} for correct factors.}
\beqn
\widetilde T_{\mu \nu} = \frac{m_{\rm p}^2}{L^2}   \sqrt{\chi}  \left( 3 \epsilon_{\mu \alpha} \epsilon^{\alpha}_{\;\; \nu} 
- 4 \epsilon_{\mu \nu} \epsilon + \chi g_{\mu \nu} \right) + \frac{\lambda^2 m_{\rm p}^2 }{L^4} \left( -4 u_{\mu} u_{\nu} + g_{\mu \nu} u^2 \right)	~.
\eeqn
Then we make an ansatz for a Friedmann-Robertson-Walker ({\sc{FRW}}) cosmology with the line-element
\beqn
{\rm d} s^2 = - {\rm d}t^2 + e^{2\nu(t)} \left( dr^2 + r^2 {\rm d} \theta^2 + r^2 \sin(\theta)^2 {\rm d}\phi^2 \right) ~.
\eeqn
The symmetry requirements of {\sc{FRW}} lead us to parameterize the displacement-vector in the form
\beqn
u_t = N L e^{2 \mu(t)}~,~u_r=u_\theta=u_\phi =0~,
\eeqn
where $N$ is a dimensionless normalization factor to be determined later. 

With these expressions, the kinetic term takes the form
\beqn
\chi = N^2 L^2 e^{4 \mu} \left(\frac{4}{3}  \dot \mu^2 +16 \dot \mu \dot \nu + 9 \dot \nu^2  \right) ~,
\eeqn
where a dot marks a derivative with respect to $t$. The non-vanishing components of $\widetilde {\bf T}$ are
\beqn
\widetilde \rho &:=& \widetilde T^0_0 = \frac{m_{\rm p}^2}{3} N^2 e^{4 \mu} \sqrt{\chi} \left( -8  \dot \mu^2  -24 \dot \mu \dot \nu + 27 \dot \nu^2 \right) - N^2 \lambda^2  \frac{m_{\rm p}^2}{L^2} e^{4\mu}~,\\
\widetilde p &:=& \widetilde T^r_r = \widetilde T^\theta_\theta = \widetilde T^\phi_\phi =
8 m_{\rm p}^2 N^2 e^{4 \mu} \sqrt{\chi} \left( \dot \mu \dot \nu +  \frac{1}{6}\dot \nu^2 \right) + N^2 \lambda^2 \frac{m_{\rm p}^2}{L^2} e^{4\mu}~,
\eeqn
the field equations read
\beqn
- m_p^2 3 \dot \nu^2 &=& T^0_0 ,~\\
- m_p^2 \left( \ddot \nu + 3 \dot \nu^2 \right) &=& T^r_r ~,
\eeqn
and the one non-vanishing equation of motion is
\beqn
9 L^2 e^{2 \mu}  \left( \frac{3}{4} \dot \nu^2 + \frac{1}{2} \dot \nu \dot \mu + \frac{1}{3} \dot \mu^2+  \ddot \nu + \frac{1}{6} \ddot \mu \right) \sqrt{\chi} = \frac{1}{2}  \lambda^2 ~.
\eeqn

To make headway on these equations, we will now assume that in case the imposter field has no mass, $\lambda =0$, the ratio
$w= \widetilde p/\widetilde \rho$ should be constant. This gives the relation
\beqn
w\left(-8 \dot \mu^2 - 24 \dot \mu \dot \nu + 27 \dot \nu^2 \right) =  4 \dot \mu \left( \dot \mu + 6 \dot \nu \right)~.
\eeqn
This equation will only be fulfilled if $\mu = C_4 \nu$, where $C_4$ is a constant, plus another constant that is irrelevant because it can be absorbed in the normalization of ${\bf u}$. 

With the, so improved, ansatz we quickly see that the particularly simple case $C_4 = 0$ (ie, ${\bf u}$ constant or $w=1$) solves both the field equations and the equations of motion
for $\nu(t)= \sqrt{\Lambda^*} t$, provided that the parameters are related as follows
\beqn
N= \frac{1}{3} \left( \frac{6}{L \sqrt{\Lambda^*}} \right)^{1/3}~,~\lambda^2 = \frac{27}{2} N \left( L \sqrt{\Lambda^*}\right)^{3} ~. \label{cosmparam}
\eeqn
In other words, the imposter field can act as a cosmological constant.

We have added an asterisk to $\Lambda^*$ because we had previously identified the cosmological constant $\Lambda=1/L^2$. 
We now see that this relation holds only up to a prefactor which depends on $\lambda$, the mass of the field. Alternatively, we
could push the prefactors into the coupling constants of the Lagrangian. We should also keep in mind, however,
that in this model the cosmological constant which we
observe is the one associated, not with ${\bf g}$, but with ${\bf{g-un}}/L$. If $\bf u$ is constant, this means we have to rescale the $t$-
coordinate to obtain the physically relevant $\Lambda$. 

An important consequence of the relations (\ref{cosmparam}) is that when we assume $\Lambda^*$ is of the same order of magnitude
as $1/L^2$, then $N\sim1$, $u_0\sim L$ and $\phi \sim 1$. This means the corrections to $g_{00}$ are of order one in the long-distance
limit, supporting the previously raised suspicion that the correction to the Newtonian potential from the integration constant in Eq. (\ref{phiapprox})
might not be negligible. 

It is worth noting that the ability of the imposter field to behave like a cosmological constant does not depend on the fractional power in the Lagrangian. It comes about because
the strain tensor  is symmetric and hence (in contrast to the field-strength tensor of a gauge-field) can have
diagonal elements, and the mass-term of the vector field makes a contribution the stress-energy of the form $u_\nu u_\mu$
which (in contrast to a scalar field) is not proportional to $g_{\mu \nu}$. For the right choice of constants, the two
contributions can combine so that the stress-energy tensor of ${\bf u}$ is proportional to the metric, which means it
behaves like a cosmological constant. This is only possible if the mass-term is included.
 
Seeing that the imposter field can mimic a cosmological constant on long distances, one can try to improve
the analysis of the previous subsection about the spherically symmetric case by studying the Schwarzschild
solution in asymptotic de-Sitter space. In this case the metric is similar to
Eq.\ (\ref{ssm}), just with $\gamma(r) = 1 - 2M/(m_{\rm p}^2 r) - \Lambda r^2$. 

One can indeed use the
method from the previous subsection to solve the equations of motion also in this background.
However, one finds that this does not reproduce the exact solution found here in the case
$M=0$. This highlights that on cosmological distances neglecting the backreaction and
mass of the field is a bad approximation. For this reason we unfortunately cannot use
the Schwarzschild metric with de-Sitter limit to fix the integration constant in Eq.\ (\ref{phiapprox}).

\section{Discussion}

Since the publication of \cite{Verlinde:2016toy}, several follow-up works have appeared \cite{Ettori:2016kll,Liu:2016nwt,Diez-Tejedor:2016fdn,Milgrom:2016huh,Brouwer:2016dvq,Hees:2017uyk,Lelli:2017sul} that
test the apparent dark matter distribution derived in \cite{Verlinde:2016toy} against data. 
Recent analyses signal a tension with data \cite{Hees:2017uyk,Lelli:2017sul}. Such conclusion, however, seems 
premature. The derivation provided in \cite{Verlinde:2016toy} uses many simplifications
and it is not at all clear in which cases these estimates are good approximations. 

We have here made progress and derived fully covariant equations. Neglecting
backreaction, we have
also obtained the solution in the background
of a Schwarzschild metric. We notice that in the covariant case the solution depends on
an integration constant which cannot be fixed in the limit that the solution has been
obtained. So long as no expression for this constant has been derived from a
suitable boundary condition, it should be treated as a free parameter.

To obtain the correct apparent dark matter distribution,
one would moreover have to solve this equation for the density profiles of galaxies and galaxy
clusters. Only with this solution can one tackle the question whether this model
fits the data. It would of course also be desirable to find a solution that includes the backreaction
from the imposter field's stress-energy on the metric. 

Furthermore, it would be of special interest to include a baryonic source in the {\sc FRW}-case, to see if
one can reproduce the relation between the cosmological constant and apparent dark
matter suggested (with reservation) in \cite{Verlinde:2016toy}. With that solution, one
could then move towards cosmological perturbation theory.

\section{Conclusion}

We have studied here a covariant version of modified gravity that has the potential to
explain dark matter and dark energy and derived the equations of motion as
well as the stress-energy-tensor. We have demonstrated that approximations
derived in the flat-space limit might not accurately describe the distribution of
the apparent dark matter. The work presented here may enable the derivation of observables
not only on galactic but also on cosmological scales.

\bigskip 
A Maple worksheet to accompany the calculations summarized here can be
downloaded at {\url{http://sabinehossenfelder.com/Physics/imposter.mw}}

\section*{Acknowledgements}

I thank Christine C.~Dantas, Stefan Scherer, and Erik Verlinde for helpful discussion. This work is supported by the Foundational Questions Institute (FQXi).

\section*{Appendix: Choice of Parameters in Lagrangian}

The canonical stress-energy
tensor of a field theory derived from Noether's law is in general not identical to the gravitational
stress-energy-tensor that acts as a source for Einstein's field equations and is derived by variation
with respect to the metric. It is a rarely noted but peculiar fact that for the fields we usually
deal with the conservation of one implies conservation of the other. In general,
however, this isn't so. 

The reason, to make a long story short, is that covariant derivatives 
don't commute. This, however, does not play a role for scalar-fields (because
the covariant derivatives reduce to partial derivatives), for gauge-bosons (because
of the anti-symmetry of the field-strength tensor) and for Dirac fields (because
the Lagrangian identically vanishes if the equation of motion is fulfilled). In other
words, this does not play a role for all fields in the Standard Model. It is hence
normally assumed in the cosmological context that the conservation of stress-energy
implies that the equations of motions are fulfilled. 

But it does play a role for the kinetic terms in the Lagrangian of the
imposter field. Even leaving aside the power of $3/2$, the equations of
motion for this field do not in general imply the conservation of the gravitational
stress-energy. This is an added complication because it means four more
equations have to be solved. 

This brings up the question whether there are any parameter combinations for
which this complication does not appear. We already know that this is so for
$\alpha=0, \beta = -\gamma$, which is the usual case for gauge-bosons. Are
there any other?

An easy way to study the role of these parameters is to put the
field in the simplest non-flat background, a de-Sitter space. Assuming that
the field is constant and its spatial components vanish, one sees that the
requirement of gravitational stress-energy-conservation in this background
is fulfilled together
with the equations of motion, if and only if $-3\alpha=4(\beta+\gamma)$. 
Assuming that $\beta=\gamma$ and further using the common normalization
$\beta=-1/2$ explains the choice we have made in this work.

\end{document}